\documentclass[11pt]{article}
\usepackage{theorem}
\usepackage{latexsym}
\usepackage{amssymb}
\usepackage{amsmath,pxfonts,color}
\newtheorem{thm}{Theorem}[section]

\newtheorem{prop}[thm]{Proposition}
\newtheorem{lem}[thm]{Lemma}

\theorembodyfont{\rm}
\newtheorem{defn}[thm]{Definition}
\newtheorem{ex}[thm]{Example}

\newtheorem{rem}[thm]{Remark}

\def\fin   {\hfill{$\Box$}\vspace{5mm}}
\def\proof {\noindent{\it Proof.}$\quad$$\quad$}
\def\l     {\left}
\def\r     {\right}
\def\<     {\langle}
\def\>     {\rangle}

\def\wt#1  {\widetilde#1 }
\def\wh#1  {\widehat#1 }
\def\ol#1  {\overline#1 }
\def\ul#1  {\underline#1 }

\def\calF  {{\cal F}}

\def\calP  {{\cal P}}
\def\calQ  {{\cal Q}}
\def\calR  {{\cal R}}

\def\VecR  {\mathbf R}
\def\ve    {\varepsilon}

\begin{document}
\title{Convex risk measures for good deal bounds}
\author{Takuji Arai\footnote{
        Department of Economics, Keio University,
        Minato-ku, Tokyo, 108-8345 Japan}
        \  and Masaaki Fukasawa\footnote{
        Department of Mathematics,
        Osaka University, 1-1 Machikaneyama, 560-0043 Japan}}
\maketitle

\begin{abstract}
We study convex risk measures describing the upper and lower bounds
of a good deal bound, which is a subinterval of a no-arbitrage pricing bound.
We call such a convex risk measure a good deal valuation and
give a set of equivalent conditions for its existence in terms of market.
A good deal valuation is characterized by  several equivalent properties and
in particular, we see that a convex risk measure is
a good deal valuation only if it is given as a risk indifference price.
An application to shortfall risk measure is given.
In addition, we show that the no-free-lunch (NFL) condition 
is equivalent to the existence of a relevant convex risk
measure which is a good deal valuation.
The relevance turns out to be a condition for 
a good deal valuation to be reasonable.
Further we investigate conditions under which any good deal
valuation is relevant. \\

\noindent
{\bf Keywords:} Convex risk measure, Good deal bound, Orlicz space,
                Risk indifference price, Fundamental theorem of asset pricing
\end{abstract}
       
\setcounter{equation}{0}
\section{Introduction}
The no-arbitrage framework in mathematical finance
is not sufficient for providing
a unique price for a given contingent claim in an incomplete market.
Instead provided is only a no-arbitrage pricing bound.
Since it is in general too wide to be useful
in financial practice, needed is an alternative way to find
nice candidates of prices of contingent claims.
As a method to give a sharper pricing bound,
the framework of no-good-deal has been discussed in much literature;
for example,
\cite{A11} \cite{B09} \cite{BL00} \cite{BS06} \cite{CGM01}
\cite{CH02} \cite{C03} \cite{CS00} \cite{JK01} \cite{KS07}
\cite{LPST05}  \cite{S04}.
The no-arbitrage pricing bound for a claim is obtained by excluding prices
which enable either a seller or buyer to enjoy an arbitrage opportunity
by trading the claim and selecting a suitable portfolio strategy.
The price in a market should be consistent with this bound 
to make the market viable.
On the other hand, an upper (resp.~a lower) good deal bound may be
interpreted as determined by
the seller's (resp.~the buyer's) attitude to the risk associated with
the claim.
This can be considered as a generalization of the both pricing
principle of no-arbitrage and exponential utility indifference valuation.
Denote by $a(x)$ such an upper bound for a claim $x$.
The functional $a$ is supposed to have the following properties:
\begin{enumerate}
\item $a(0) = 0$,
\item $a(x)\leq a(y)$ if $x\leq y$,
\item $a(x+c)=a(x)+c$ for any $c\in \mathbb{R}$,
\item $a(\lambda x+(1-\lambda)y)\leq\lambda a(x)+(1-\lambda)a(y)$
      for any $\lambda\in[0,1]$
\end{enumerate}
for any claims $x$ and $y$.
In the second property, the inequality $x \leq y$ is in the almost sure sense,
where we regard the claims as random variables. 
In the third, the element $c \in \mathbb{R}$ stands for
a deterministic cash-flow.
The last one represents the risk-aversion of the seller
taking into account the impact of diversification.
In brief, we suppose that $\rho_a$ defined as $\rho_a(x):=a(-x)$
is a normalized convex risk measure.
If we impose additionally the positive homogeneity:
$a(\lambda x) = \lambda a (x)$ for all $x$ and $\lambda \geq 0$,
which implies the subadditivity:
$a(x+y) \leq a(x) + a(y)$ for all $x$ and $y$,
then $\rho_a$ becomes a coherent risk measure.
By the same sort argument as above, a functional $b$ which refers to a
lower good deal bound is given by a normalized convex risk measure
$\rho_b$ as $b(x)=-\rho_b(x)$.

A good deal bound should be a subinterval of the no-arbitrage pricing bound,
so not every convex risk measure yields a good deal bound. 
The aim of this paper is to characterize such a convex risk measure,
which we call a good deal valuation (GDV hereafter); we 
define GDV  as a normalized convex risk measure $\rho$ with the Fatou property
such  that for any claim $x$,
the value $\rho(-x)$ lies in the no-arbitrage pricing bound of $x$.
This definition of GDV is given from sellers' viewpoint;
for a GDV $\rho$ and a claim $x$, 
$a(x):= \rho(-x)$ serves as an ask price of $x$.
Nevertheless, it is easy to see that if $\rho$ is a GDV, then
$b := -\rho$ gives bid prices.
We impose the Fatou property as a natural continuity condition for good
deal bounds. 

First we investigate equivalent conditions for the existence of a GDV.
Among others, we show that a  GDV exists under a condition
weaker than the no-arbitrage one, which means that there may be
GDVs even if the underlying market admits an arbitrage opportunity.
Further we study equivalent conditions for a given $\rho$ to be a GDV.
In particular, we see that any GDV is given as a risk indifference price.
The concept of risk indifference price has been undertaken by \cite{X06}.
There is much literature on this topic
(\cite{ES10} \cite{KS07-2} \cite{OS09} among others).
Some of the above papers observe that a risk indifference price provides
a good deal bound.
Our assertion is that its reverse implication also holds true, 
which seems a new insight.

As mentioned before, GDV may exist even in markets with free lunch.
We observe the equivalence between the no-free-lunch condition (NFL) 
and the existence of a relevant GDV, that is
a relevant convex risk measure which is a GDV.
This could be considered as a version of Fundamental Theorem of
Asset Pricing (FTAP).
Moreover as a version of Extension Theorem, we see that
the relevance of a GDV is equivalent to that
the extended market by the GDV satisfies NFL.
We see also that the relevance is equivalent to the no-near-arbitrage
condition (NNA) introduced by \cite{S04}.
We give an example (Example~\ref{ex5-1})
which shows that NFL for the original market
does not ensure NNA in general for a given GDV. 
We investigate conditions under which any GDV is relevant,
and illustrate some examples related to this topic.

Now we mention the preceding results on FTAP from the viewpoint of
good deal bound.
Kreps \cite{K81} introduced  NFL  and proved FTAP as well as Extension Theorem.
\v{C}ern\'y and Hodges \cite{CH02} established the 
framework of good deal bound and gave a version of Extension Theorem.
Jaschke and K\"uchler \cite{JK01} 
showed that good deal bounds are essentially equivalent to
coherent risk measures and gave a variant of FTAP.
Staum \cite{S04} extended their results to the noncoherent case.
Bion-Nadal~\cite{B09} introduced a dynamic version and gave an
associated FTAP.
In \cite{JK01} and \cite{S04}, an acceptance set reflecting an investor's
preference is given first, and a convex risk measure induced by it
is considered as a functional describing a good deal bound.
Our approach is different, although we treat very similar problems.
In our study, a convex risk measure is given first,
and necessary and sufficient conditions for the given convex risk measure
to be a GDV is discussed.
This approach is in the same spirit as \cite{B09}. 
Our results provide a deeper understanding of a convex risk measure
as a pricing functional in a market.
Although our framework appears to be static, an extension to the dynamic
framework of \cite{B09} can be done in a straightforward manner.
A detailed comparison with \cite{S04} and \cite{B09} will be given in
Remarks~\ref{rem-thm2-2}, \ref{staumrem} and \ref{bionnadalrem}.

In Section 2, we describe our model and prepare notation.
In particular, we introduce the definitions and some basic properties of
superhedging cost and risk indifference price.
Main results are given in Sections 3 and 4.

\setcounter{equation}{0}
\section{Preliminaries}
Here we introduce our framework and several  basic results.

\subsection{The Orlicz space}
Let $(\Omega, \calF, \mathbb{P})$ be a complete probability space.
The Orlicz space $L^{\Psi}$ with Young function $\Psi$ is defined as
the set of the random variables $X$ such that there exists $c > 0$,
\begin{equation*}
\mathbb{E}[\Psi(cX)] < \infty.
\end{equation*}
Here we call 
$\Psi : \mathbb{R} \to \mathbb{R}\cup \{\infty\}$ 
a Young function if it is an even convex function with 
$\Psi(0)=0$, $\Psi (x) \uparrow \infty$ as $x \uparrow \infty$ and
$\Psi(x) < \infty$ for $x$ in a neighborhood  of $0$.
It is a Banach lattice with the gauge norm
\begin{equation*}
\|X\|:= \inf\{c > 0; \mathbb{E}[\Psi(X/c)] \leq 1 \}
\end{equation*}
and pointwise ordering in the almost sure sense.
In the case of $\Psi = \Psi_{\infty}$:
\begin{equation*}
\Psi_{\infty}(x) := \begin{cases} 0  & \text{ if } |x| \leq 1, \\
\infty &  \text{ otherwise }
\end{cases}
\end{equation*}
we have $L^\Psi = L^\infty$. Further, for $\Psi_p(x) := |x|^p$ with
$p\geq 1$, we have $L^{\Psi_p} = L^p$.
The Orlicz heart $M^\Psi$ is a subspace of $L^\Psi$ defined as
\begin{equation*}
M^\Psi := \{ X \in L^\Psi| \mathbb{E}[\Psi(cX)] < \infty \text{ for all }
c > 0\}.
\end{equation*}
In this paper we consider the set of the future cash-flows $L$ to be either
$L^\Psi$ or $M^\Psi$
with a fixed Young function $\Psi$. 
This specification would be justified by noting that $L$ becomes 
a linear space of random variables with natural ordering 
and sufficiently abstract in that
it incorporates $L^p$ spaces with $1 \leq p \leq \infty$.
More importantly, a Young function $\Psi$ may be connected to a utility
function $u$ as $\Psi(x) = -u(-|x|))$ 
and then $L$ becomes a suitable space where expected utility
maximization is considered (see e.g., \cite{BF09}).
Note that the case of exponential utility is covered.
Our treatment and results do not depend on a specific choice of $\Psi$.
This generality is indeed necessary to derive a conclusion which does
not depend on a specific choice of utility function.

Let $M \subset L$  be the set of the $0$-attainable claims.
Each element of $M$ represents a future payoff
which investors can super-replicate with $0$ initial endowment.
Simultaneously, $M$ might be regarded as the set of strategies which
investors can take.
We suppose that $M$ is a convex cone including $L_-$,
where we denote $L_+$ (resp. $L_-$)$:=\{x\in L|x\geq0$ (resp. $\leq$)$\}$.

Let $L^\ast_+$ be the set of all positive linear functionals on $L$.
Remark that any element of $L^\ast_+$ is continuous by the
Namioka-Klee theorem (see \cite{BF09} for an extended result).
The both cases of $L=L^\Psi$ and $L=M^\Psi$
are  treated in a unified way in the following.
Let $L^\dagger := L^{\Psi^\dagger}$, where $\Psi^\dagger$  is 
the complimentary function of $\Psi$ defined as
\begin{equation*}
\Psi^\dagger(y) := \sup_{x \in \mathbb{R}}\{xy-\Psi(x)\}.
\end{equation*}
Define a set of probability measures
$\calP :=\{Q \ll \mathbb{P}|\mathrm{d}Q/\mathrm{d}\mathbb{P}\in L^\dagger\}$.
Further, let $\ol{L^*} :=\l\{g \in L^*_+ | g(1)=1,
g(m) \leq0\mbox{ for any }m\in M\r\}$,
$\calQ :=\{Q\in\calP| \mathrm{d}Q/\mathrm{d}\mathbb{P}\in\ol{L}^*\}$, 
and $\calQ^e := \{Q\in\calQ|Q\sim \mathbb{P}\}$.
For $Q \in \calP$, denote by $\mathbb{E}_Q$ the corresponding
expectation operator.
By Young's inequality:
\begin{equation*}
\frac{xy}{ab} \leq \Psi(\frac{x}{a}) + \Psi^\dagger(\frac{y}{b})
\end{equation*}
for any $x, y \in \mathbb{R}$ and $a,b >0$,
the operation $\mathbb{E}_Q$ enables us to identify $\calP$ with a subset of 
$L^\ast_+$.

\subsection{Convex risk measure}
Here we collect several notions and 
results on convex risk measures which we utilize in this paper.
A convex risk measure $\rho$ is
a $(-\infty,+\infty]$-valued functional on $L$ satisfying
\begin{description}
\item[properness:]      $\rho(0) < \infty$,
\item[monotonicity:]    $\rho(x)\geq\rho(y)$ 
                        if $x\leq y$,
\item[cash-invariance:] $\rho(x+c)=\rho(x)-c$
                        for any $c\in\mathbb{R}$,
\item[convexity:]       $\rho(\lambda x+(1-\lambda)y)
                        \leq\lambda\rho(x)+(1-\lambda)\rho(x)$
                        for  any $\lambda\in[0,1]$,
\end{description}
for any $x$, $y\in L$.
A convex risk measure $\rho$ is a 
{\bf coherent risk measure} if it satisfies in addition,
\begin{description}
\item[positive homogeneity:]
                        $\rho(cx)=c\rho(x)$ for any $x\in L$ and any $c>0$.
\end{description}

\begin{thm}[Biagini and Frittelli \cite{BF09}] \label{repfa}
Let $\rho$ be a convex risk measure. Then,
\begin{equation*}
\rho(-x) = \max_{g \in L^\ast_+, g(1)=1}\{g(x) - \rho^*(g)\}
\end{equation*}
for $x \in \mathrm{Int}\{\rho < \infty\}$, where for $g \in L^{\ast}_+$,
\begin{equation*}
\rho^*(g) := \sup_{x \in L}\{ g(x) - \rho(-x)\}.
\end{equation*}
\end{thm}

\noindent
A convex risk measure $\rho$ is said to have {\bf the Fatou property} if
for any increasing sequence $\{x_n\} \subset L $ with
$x_n \uparrow x_{\infty}$ a.s., $\rho(-x_n) \uparrow \rho(-x_{\infty})$. 
Denote by $\calR$ the set of all convex risk measures
with $\rho(0)=0$ and the Fatou property.

\begin{thm}[Biagini and Frittelli \cite{BF09}]\label{BF}
For $\rho \in \calR$, we have for $x \in L$,
\begin{equation}
\label{eq-repre1}
\rho(x)=\sup_{Q \in\calP}\l\{ \mathbb{E}_Q[-x]-\rho^{\ast}(Q)\r\}.
\end{equation}
\end{thm}
A convex risk measure $\rho$ is said to be {\bf finite} if
$\rho(x) < \infty$ for all $x \in L$.
\begin{rem}
In the case of $L=M^\Psi$,
it is known that $L^\dagger$ coincides with the dual of $L$ and
the supremum in (\ref{eq-repre1}) is attained.
Moreover, every finite convex risk measure has the Fatou property.
See \cite{BF09} for the detail.
The finiteness condition cannot be dropped as we see in
Example~\ref{ex1} below.
If
$\Psi$ satisfies the $\Delta_2$ condition: there exist $t_0 > 0$ and 
$K > 0$ such that
$\Psi(2t) \leq K \Psi(t)$ for any  $t \geq t_0$,
then we have $L^\Psi = M^\Psi$.
For $ p \in [1,\infty)$, $L^p$ is an example of such cases.
\fin
\end{rem}

\noindent
A convex risk measure $\rho$ is said to have {\bf the Lebesgue property} if
for any sequence $\{x_n\} \subset L$ 
with $\sup_n\|x_n\|_{\infty} < \infty $ and 
$x_n \to x_{\infty}$ a.s., it holds that $\rho(x_n) \to \rho(x_{\infty})$ as 
$n \to \infty$.
Here $\|\cdot\|_{\infty}$ refers to the $L^\infty$ norm.
This definition was introduced in \cite{JST} for the $L=L^\infty$ case.
Since any continuous linear
functional on L can be decomposed into the sum of an element of 
$L^\dagger \subset L^1$
and a purely finitely additive signed measure (see [23]), 
the same argument as the proof of Theorem~2.4 in \cite{JST}
can apply to have the following result
with the aid of Theorem~\ref{repfa} above.

\begin{thm}\label{Leb}
For a finite convex risk measure $\rho$, the following are equivalent:
\begin{enumerate}
\item $\rho$ has the Lebesgue property.
\item for any $\alpha > 0$ and a sequence of measurable sets $A_n$ with
      $P(A_n) \to 0$, it holds that
      $\rho(-\alpha 1_{A_n}) \to 0$    as $n \to \infty$.
\item for any $c > 0$, the set $\{g \in L^{\ast}_+; \rho^*(g) \leq c\}$ is a
      uniformly integrable subset of $L^\dagger$
      and for any $x \in L$, it holds that
\begin{equation}\label{repL}
\rho(-x) = \max_{Q \in \calP}\{\mathbb{E}_Q[x]-\rho^\ast(Q)\}.
\end{equation}
\end{enumerate}
\end{thm}

\noindent
Note that the Fatou property follows from the Lebesgue property
by (\ref{repL}).\\

\noindent
A convex risk measure is said to be {\bf relevant } if $\rho(-z) > 0$
for any $z \in L_+\setminus \{0\}$.
The relevance was introduced in \cite{Delrel} as 
a condition for coherent risk measures with the Fatou property to be
represented as (\ref{eq-repre1}) with a set of equivalent probability
measures instead of $\calP$.\\

\subsection{Superhedging cost}
Here we discuss superhedging cost.
Define a functional $\rho^0$ on $L$ as
\begin{equation}
\label{eq-rho0}
\rho^0(x):=\inf\{c\in\mathbb{R}|\mbox{ there exists $m\in M$ such that }
           c+m+x\geq0\}.
\end{equation}
Since $\rho^0(-x)$ represents the superhedging cost for a claim $x$,
it gives the upper no-arbitrage pricing bound for $x$. In fact
if a seller could sell $x$ with a price greater than $\rho^0(-x)$,
then she could enjoy an arbitrage opportunity by taking a suitable
strategy from $M$.
By the same reasoning  the lower no-arbitrage pricing bound for $x$
is given by $-\rho^0(x)$.


\begin{lem}
\label{lem0}
The superhedging cost $\rho^0$ is $(-\infty,\infty]$-valued if and only
 if  $\ol{L}^*\neq\emptyset$.
If $\rho^0$ is $(-\infty,\infty]$-valued, then it is a coherent risk
 measure with
\begin{equation*}
(\rho^0)^\ast(g) =
\begin{cases} 
      0      & \text{ if } g \in \ol{L}^{\ast}, \\
      \infty & \text{otherwise.}
\end{cases}
\end{equation*}
\end{lem}
\proof
Suppose that  $\ol{L}^*\neq\emptyset$.
If there exists $x\in L$ with $\rho^0(x)=-\infty$, then
(\ref{eq-rho0}) implies that for any $c>0$, we can find  $m^c\in M$
such that $-c+m^c+x\geq0$.
This gives $g(x) \geq c$, so that $g(x) = \infty$ for any  $g\in\ol{L}^*$.
This is a contradiction, so  $\rho^0$ is $(-\infty, \infty]$-valued.
Next, suppose that  $\ol{L}^\ast = \emptyset$. Then there exists a sequence 
$\{m_n\} \subset M$ such that $\|m_n-1\| \to 0$ as $n \to \infty$.
In fact if the closure $M^s$ of $M$ does not include $1$, then 
the Hahn-Banach theorem implies the existence of a continuous linear
functional $\mu$ such that  $\mu(1) > \sup_{m \in M^s} \mu(m)$.
The RHS is $0$ since $M^s$ is a cone.
That $L_- \subset M^s$ implies $\mu \in L^\ast_+$. 
This means $\ol{L}^\ast \neq \emptyset$, which is a contradiction.
Now, taking a subsequence if necessary, we may suppose that 
$\sum_{n =1}^\infty\|m_n-1\| < \infty$.
Then for $x := \sum_{n=1}^{\infty}|m_n-1| \in L$ and for 
all $N \in \mathbb{N}$,
\begin{equation*}
x \geq  \sum_{n=1}^N (1-m_n) = N - \sum_{n=1}^Nm_n,
\end{equation*}
which implies that $\rho^0(x)\leq -N$, and so $\rho^0(x) = -\infty$.

Now we see that $\rho^0$ is a coherent risk measure and
calculate $(\rho^0)^\ast$.
The convexity and positive homogeneity of $\rho^0$ 
follow from the assumption that  $M$ is a convex cone.
The monotonicity and cash-invariance are obvious.
The fact that $\rho^0(0)\leq0$ implies that $(\rho^0)^\ast(g) \geq 0$
for any $g \in L^\ast_+$.
On the other hand, for any $\ve>0$ and $x \in L$, 
we can find $m^\ve\in M$ so that $\rho^0(x)+\ve+m^\ve+x\geq0$.
Since $g(m^\ve)\leq0$ for $g \in \ol{L}^\ast$, we have
$\rho^0(x)+\ve\geq g(-x)$, which implies that
\[
\sup_{x\in L}\{g(-x)-\rho^0(x)\}\leq0.
\]
We therefore have $(\rho^0)^\ast(g)=0$ for $g \in \ol{L}^\ast$.
For $g\in L^\ast_+ \setminus\ol{L}^\ast$,
there exists  $m\in M$ such that $g(m)>0$.
Since $M$ is a cone,
\[(\rho^0)^\ast(g) = 
\sup_{x\in L}\{g(-x)-\rho^0(x)\}
              \geq\sup_{m\in M}\{g(m)-\rho^0(-m)\}
              \geq\sup_{m\in M}g(m)=\infty.
\]
\fin

\noindent
For later use, we define for $x \in L$,
\begin{equation*}
\wh{\rho^0}(x) :=
\begin{cases}
      \sup_{Q \in \calQ} \mathbb{E}_Q[-x] & \text{ if } \calQ \neq \emptyset \\
      -\infty                             & \text{ otherwise.}
\end{cases}
\end{equation*}
By definition $\wh{\rho^0}$ is a coherent risk measure on $L$ belonging to
$\calR$ if $\calQ \neq \emptyset$.

\begin{lem}
\label{lem0-2}
If $\calQ \neq \emptyset$,
then
$-\rho^0(x) \leq -\wh{\rho^0}(x) \leq 
\wh{\rho^0}(-x) \leq \rho^0(-x)$ for any $x \in L$.
Moreover if $\calQ^e \neq \emptyset$, then  $\wh{\rho^0}$ is relevant. 
\end{lem}

\proof
For any $x \in L$ and $\ve>0$, there exists  $m\in M$ such that
$\rho^0(x)+\ve+m+x\geq0$.
Then we have
$\mathbb{E}_Q[-x] \leq\rho^0(x)+\ve$
for any $Q\in\calQ$.
Since $Q\in\calQ$ and $\ve>0$ are arbitrary,
we have $\wh{\rho^0}(x)\leq\rho^0(x)$.
It suffices then to observe that $\wh{\rho^0}(x) + \wh{\rho^0}(-x) \geq 2
\wh{\rho^0}(0)=0$ by the convexity.

The relevance under $\calQ^e \neq \emptyset$ is shown by noting that
\begin{equation*}
\wh{\rho^0}(-x) = \sup_{Q \in \calQ}\mathbb{E}_Q[x] = 
\sup_{Q \in \calQ^e}\mathbb{E}_Q[x].
\end{equation*}
In fact if there exists $Q_1 \in \calQ$ with
$\mathbb{E}_{Q_1}[x] > \sup_{Q \in \calQ^e} \mathbb{E}_Q[x]$,
then  we have a contradiction since
for any $Q_0 \in \calQ^e$,
$\lambda Q_0 + (1-\lambda)Q_1 \in \calQ^e$ converges to  $Q_1$
in $\sigma(L^\dagger,L)$ as $\lambda \downarrow 0$.
\fin

\noindent
The following example shows that
$\rho^0$ does not necessarily coincides with $\wh{\rho^0}$,
so is not always represented as (\ref{eq-repre1})
even though $\calQ$ is not empty.

\begin{ex}
\label{ex1}
Let $L =L^p$ with $p \in [1,\infty)$ and take the following set as $M$:
\[
M=\{-z+ \mathbb{E}_{Q^0}[z]|z\in L_+\}-L_+,
\]
where $Q^0 \in \calP$ is arbitrarily fixed.
Any element of $M$ is bounded from above.
Therefore by the definition of $\rho^0$, we have 
$\rho^0(-z)=\infty$ for $z\in L_+$ which is not bounded from above.
It is clear that $Q^0\in\calQ$, so that
$\ol{L}^\ast \neq \emptyset$.
Therefore $\rho^0$ is a coherent risk measure by Lemma \ref{lem0}.
Moreover  $\calQ =\{Q^0\}$ since for any  $Q \in \calQ$,
we have  
$\mathbb{E}_{Q^0}[z]\leq \mathbb{E}_Q[z]$ for any $z\in L_+$,
which implies that $Q=Q^0$.
Therefore $\rho^0$ cannot be represented as (\ref{eq-repre1}).

In fact we can prove that $\rho^0$ does not have the Fatou property.
Let $z\in L_+$ be unbounded from above.
Consider the increasing sequence  $z_n = z\wedge n$, $n \in \mathbb{N}$.
Since $n-z_n\in L_+$, we have $z_n-\mathbb{E}_{Q^0}[z_n] \in M$.
It follows that  $\rho^0(-z_n)\leq \mathbb{E}_{Q^0}[z_n]\to E_{Q^0}[z]<\infty$,
while $\rho^0(-z) = \infty$.
\fin
\end{ex}

\subsection{Risk indifference prices}
Here we recall risk indifference price.
Given a convex risk measure $\rho$, define a functional  $I(\rho)$  on $L$ as
\begin{eqnarray} \label{eq-Ieta}
I(\rho)(x)
&:=\inf\l\{c\in\mathbb{R}|\inf_{m\in M}\rho(c+m+x)\leq\inf_{m\in M}\rho(m)\r\}
   \nonumber \\
&= \inf\l\{c\in\mathbb{R}|\inf_{m\in M}\rho(m+x)-c\leq\inf_{m\in M}\rho(m)\r\}.
\end{eqnarray}
Then $I(\rho)(-x)$ describes the risk indifference seller's price for $x$
induced by $\rho$ as introduced in \cite{X06}.
The idea is explained as follows.
If a trader sells a claim $x$ with a price $c>I(\rho)(-x)$,
then she can find  $\wh{m}\in M$ such that
$\rho(c+\wh{m}-x)\leq\inf_{m\in M}\rho(m)$.
This means that selling the claim with the price does not increase the risk
measured by $\rho$.
The following lemma gives a representation of $I(\rho)$.
Denote  $\check{\rho} :=\rho -\inf_{m\in M}\rho(m)$.

\begin{lem} \label{implicit}
Let $\rho$ be a convex risk measure. 
If $I(\rho)$ is $(-\infty, \infty]$-valued,
then  we have $\inf_{m \in M}\rho(m) \in \mathbb{R}$ and  that
$I(\rho)$ is a convex risk measure with
\begin{equation*}
I(\rho)^\ast(g) 
=
\begin{cases}
\check{\rho}^\ast(g) 
=  \rho^\ast(g) + \inf_{m \in M}\rho(m), & \text{ if }
g \in \ol{L}^\ast \\
\infty & \text{ otherwise.}
\end{cases}
\end{equation*}
If $I(\rho)\in\calR$ in addition, then $\calQ \neq \emptyset$ and
      \begin{equation}
      \label{eqIeta}
      I(\rho)(x)=\sup_{Q\in\calQ}\{\mathbb{E}_Q[-x]-\check{\rho}^\ast(Q)\}.
      \end{equation}
\end{lem}

\proof
Since $\rho(0)<\infty$ and $0 \in M$,
we have $I(\rho)(0) = 0$ or $-\infty$ depending on whether
$\inf_{m\in M}\rho(m)$ is finite or $-\infty$. Therefore if 
$I(\rho) > - \infty$ then $\inf_{m\in M}\rho(m)$ is finite and
$I(\rho)(x)=\inf_{m\in M}\rho(x+m)-\inf_{m\in M}\rho(m) 
= \inf_{m \in M}\check{\rho}(x+m)$.
From this the cash-invariance and monotonicity of $I(\rho)$ 
are obvious. The convexity follows from that $M$ is convex.
Since $M$ is a cone, we have
\begin{equation*}
\begin{split}
I(\rho)^\ast(g)
=& \sup_{x \in L}\{g(-x) - I(\rho)(x)\} \\
=& \sup_{m \in M}\sup_{x \in L} \{g(-x) - \check{\rho}(x+m))\} \\
=& \sup_{m \in M}\{ g(m) + \check{\rho}^\ast(g)\} \\
=& \l\{
   \begin{array}{ll}
         \check{\rho}^\ast(g) & \mbox{if }g \in\ol{L}^\ast \\
         \infty                  & \mbox{otherwise}.
   \end{array}
   \r.
\end{split}
\end{equation*}
By Theorem~\ref{BF}, we have (\ref{eqIeta})
if $I(\rho) \in \calR$ and in particular, $\calQ \neq \emptyset$.
\fin

\setcounter{equation}{0}
\section{Good deal valuations}
In this section we discuss conditions under which a convex risk measure
yields a good deal bound.
A good deal bound should be a subinterval of the no-arbitrage pricing
bound. We therefore introduce the following definition.

\begin{defn}
A convex risk measure $\rho\in\calR$ is said to be a good deal valuation (GDV)
if
\begin{equation}
\label{eqGDV}
\rho(-x)\in[-\rho^0(x), \rho^0(-x)]\mbox{  for any }x\in L.
\end{equation}
\end{defn}

\noindent
As mentioned in Introduction,
the above definition is given from  seller's viewpoint.
Nevertheless, (\ref{eqGDV}) is equivalent to
\begin{equation}
-\rho(x)\in[-\rho^0(x), \rho^0(-x)]\mbox{  for any }x\in L,
\end{equation}
which is from buyer's viewpoint.
In addition, $-\rho(x)\leq\rho(-x)$ for any $x\in L$ because
$\rho(x)+ \rho(-x)\geq 2 \rho(0)=0$ by the convexity.
For a GDV $\rho$, a good deal bound may be constructed as
$[-\rho(x), \rho(-x)]$, which is a subinterval of $[-\rho^0(x), \rho^0(-x)]$.
Note that  the upper and lower bounds of a good deal bound may be
described by different GDVs.

\subsection{Existence of good deal valuations}
Here we present a set of equivalent  conditions for the existence of a GDV.
Denote by $\overline{M}$ the closure of $M$ in $\sigma(L,L^\dagger)$.

\begin{thm}
\label{thm1}
The following are equivalent:
\begin{enumerate}
\item $\calQ\neq\emptyset$.
\item There exists a GDV.
\item $\mathbb{P}(m>0)<1$ for any $m\in \ol{M}$.
\item $1 \notin \  \ol{M}$.
\end{enumerate}
\end{thm}

\proof
1$\Rightarrow$2: This is from Lemma~\ref{lem0-2}.

\noindent
2$\Rightarrow$1:
Let $\rho$ be a GDV.
Since $\rho(-m)\leq\rho^0(-m)\leq0$ for any $m\in M$,
\[
\rho^\ast(Q)=\sup_{x\in L}\{\mathbb{E}_Q[-x]-\rho(x)\}
         \geq\sup_{m\in M}\{\mathbb{E}_Q[m]-\rho(-m)\}
         \geq\sup_{m\in M}\mathbb{E}_Q[m].
\]
Then the cone property of $M$ implies that $\rho^\ast(Q)=+\infty$
for any $Q\in\calP\backslash\calQ$.
If $\calQ$ is empty, then $\rho$ equals to $-\infty$ identically by
(\ref{eq-repre1}), which contradicts $\rho \in \calR$.

\noindent
1$\Rightarrow$3: 
If there exists $m\in \ol{M}$ such that $P(m>0)=1$, then
we have $\mathbb{E}_Q[m]>0$ for any $Q\in\calP$, and so
$\calQ=\emptyset$.
 
\noindent
3$\Rightarrow$4: This holds true clearly.

\noindent
4$\Rightarrow$1:
Since $1 \notin \ol{M}$, 
the Hahn-Banach theorem implies that there exists $z\in L^\dagger$ such that
\begin{equation}
\label{eqHB}
\sup_{m\in \ol{M} }\mathbb{E}[zm]< \mathbb{E}[z].
\end{equation}
We have 
$\sup_{m\in\overline{M}}\mathbb{E}[zm]=0$ 
because $ 0 \in M$ and  $\ol{M}$ is a cone.
Since $L_-\subset \ol{M}$, we have then that $z \in L^\ast_+ \cap L^\dagger$,
so that $z/\mathbb{E}[z] \in \calQ$.
\fin

\noindent
Condition~3 in the above theorem is weaker than the no-arbitrage condition.
This means that a GDV may exist even if there is an arbitrage opportunity.
The following example shows that we cannot replace $\ol{M}$ with $M$
in Conditions 3 and 4.

\begin{ex}
We take the Lebesgue measure space on $(0,1]$ as the underlying probability
space $(\Omega,\calF,\mathbb{P})$.
Let $u$ be the random variable given by $u(\omega):=\omega$, and
$M$ be given by $\{cu|c\geq0\}-L_+$.
We can see several interesting facts on this example as follows:
\begin{enumerate}
\item We consider the following two conditions:
      \begin{enumerate}
      \item $\mathbb{P}(m>0)<1$ for any $m\in M$,
      \item $1 \notin M$.
      \end{enumerate}
      This example satisfies (b), but does not satisfy (a).
      Replacing $M$ by $\ol{M}$,
      the two conditions become equivalent by Theorem~\ref{thm1}.
\item Since $1 \notin M$, we have $\rho^0(0)=0$.
      Therefore if we take $L = L^\infty$,
      then $\rho^0$ is a finite coherent risk measure. 
      In fact for any $x \in L^\infty$, 
      $-\|x\|_{\infty} = \rho^0(\|x\|_{\infty}) \leq \rho^0(x) \leq
      \rho^0(-\|x\|_{\infty}) = \|x\|_{\infty}$ by monotonicity.
On the other hand,
      $\rho^0$ is not a convex risk measure on $L=L^p$ with $p \in [1,\infty)$
      since $\ol{L}^* = \calQ$ is empty.
      Note that for $x(\omega):=\log\omega$, we have $\rho^0(-x)=-\infty$.
\item Notice that $\calQ$ is empty
      despite that the above Condition (b) holds.
      We therefore need to take the closure of $M$ in Condition 4
      of Theorem \ref{thm1}.
      In fact, considering the sequence $m_n:=(nu)\wedge1$,
      $m_n$ converges to $1$, and so
      this example does not satisfy Conditions 3 nor 4.
\end{enumerate}
\fin
\end{ex}

\subsection{Equivalent conditions for good deal valuations}
Here we present conditions for a given $\rho$ to be a GDV.
The main contribution of the following theorem,
is to show the equivalence between GDVs and risk indifference prices.

\begin{thm}
\label{thm2}
For any $\rho\in\calR$, the following conditions are equivalent:
\begin{enumerate}
\item $\rho$ is a GDV.
\item $\rho(-m)\leq0$ for any $m\in M$.
\item There exists a function $c : \calQ \to \mathbb{R}$ such that
      for any $x \in L$,
      $$\rho(x)=\sup_{Q\in\calQ}\{\mathbb{E}_Q[-x]-c(Q)\}.$$ 
\item There exists  $\eta\in\calR$ such that $\rho=I(\eta)$.
\item[4$^\prime$.] $\rho=I(\rho)$, that is, $\rho$ is a fixed point of $I$.
\item $\rho(-x) \in [-\wh{\rho^0}(x),\wh{\rho^0}(-x)]$ for any $x \in L$.
\item $\{\rho^0 \leq 0\} \subset \{\rho \leq 0\}$.
\item $\calQ\supset \{Q\in\calP|\rho^\ast(Q)<+\infty\}$.
\item There exists a convex set $A \subset L$ including
      $0$ with $A+L_+\subset A$ and
      $A\cap\mathbb{R}=\mathbb{R}_+$ such that for any $x \in L$,
      \begin{equation}
      \label{eqrhoCA}
      \rho(x)=\inf\{c\in\mathbb{R}|\mbox{ there exists  $m\in M$
      such that }c+m+x\in A\}.
      \end{equation}
\end{enumerate}
\end{thm}

\proof
1$\Rightarrow $2: This is because $\rho(-m) \leq \rho^0(-m)\leq 0$
for any $m \in M$ by the definitions of GDV and $\rho^0$. 

\noindent
2$\Rightarrow$7:
We have
\[
\rho^\ast(Q)=\sup_{x\in L}\{\mathbb{E}_Q[-x]-\rho(x)\}
         \geq\sup_{m\in M}\{\mathbb{E}_Q[m]-\rho(-m)\}
         \geq\sup_{m\in M}\mathbb{E}_Q[m].
\]
Since $M$ is a cone, we have 
$\rho^\ast(Q)=\infty$ for any $Q\in \calP\setminus\calQ$.

\noindent
7$\Rightarrow$3: This is from Theorem~\ref{BF}.

\noindent
3$\Rightarrow$4$^\prime$ and 4:
Since $\rho \in \calR$, we have
\[
\rho(-m)=\sup_{Q\in\calQ}\{\mathbb{E}_Q[m]-c(Q)\}
        \leq-\inf_{Q\in\calQ}c(Q)= \rho(0)=0
\]
for any $m\in M$.
Then, by the convexity, 
we have $\rho(m) + \rho(-m) \geq 2\rho(0)=0$ and so,
$\inf_{m \in M }\rho(m) = 0$.
Therefore,
\begin{equation}
\label{eq3to4}
I(\rho)(x)=\inf_{m\in M}\rho(m+x)-\inf_{m\in M}\rho(m) \leq \rho(x)
\end{equation}
and
\begin{equation*}
I(\rho)(x) = \inf_{m \in M}\sup_{Q \in \calQ}\{ \mathbb{E}_Q[-m-x] -c(Q) \}
\geq \sup_{Q\in\calQ}\{\mathbb{E}_Q[-x]-c(Q)\} = \rho(x).
\end{equation*}

\noindent
4$\Rightarrow $5: By Lemma~\ref{implicit},
$\rho=I(\eta)$ is represented as
\[
\rho(x)=\sup_{Q\in\calQ}\l\{\mathbb{E}_Q[-x]-\check{\eta}^\ast(Q)\r\}.
\]
Since $\rho(0) = 0$, we have $\check{\eta}^\ast(Q) \geq 0$.
Therefore,
\[
\wh{\rho^0}(-x)
   =   \sup_{Q\in\calQ}\mathbb{E}_Q[x]
   \geq\sup_{Q\in\calQ}\l\{\mathbb{E}_Q[x]-\check{\eta}^\ast(Q)\r\}
   =   \rho(-x)
\]
for all $x \in L$.
It suffices then to recall that $\rho(x) + \rho(-x) \geq 2\rho(0) = 0$
by the convexity.

\noindent 5$\Rightarrow$1:
This is from Lemma~\ref{lem0-2}.

\noindent
3$\Rightarrow$6:
For any $x\in\{\rho^0\leq0\}$, Lemma~\ref{lem0-2} implies that
$\sup_{Q\in\calQ}\mathbb{E}_Q[-x]=\wh{\rho^0}(x)\leq0$.
We have then
\[
\rho(x)=   \sup_{Q\in\calQ}\{\mathbb{E}_Q[-x]-c(Q)\}
       \leq\sup_{Q\in\calQ}\{-c(Q)\}=\rho(0)=0.
\]

\noindent
6$\Rightarrow$2:
This is because $\rho^0(-m)\leq 0$ by definition.

\noindent
4$^{\prime} \Rightarrow$8:
Taking $A = \{\rho \leq 0\}$ and noting that $\inf_{m\in M}\rho(m)=0$, we have
\begin{eqnarray*}
\rho(x)&=&   I(\rho)(x)=\inf_{m\in M}\rho(m+x)
             =\inf\{c\in\VecR|\inf_{m\in M}\rho(m+x)\leq c\} \\
       &\leq&\inf\{c\in\VecR|\mbox{ there exists $m\in M$
                   such that }\rho(m+x)\leq c\} \\
       &=&   \inf\{c\in\VecR|\mbox{ there exists $m\in M$
                   such that }c+m+x\in A\} \\
       &\leq&\inf\{c\in\VecR| c+x\in A\}=\rho(x).
\end{eqnarray*}

\noindent
8$\Rightarrow$2: This is obvious.
\fin

\begin{rem}
\label{rem-thm2-2}
Denote by $\rho_A$ the RHS of (\ref{eqrhoCA}).
In \cite{JK01} and \cite{S04}, 
the set $A$ is given as an acceptance set and
$\rho_A$ is considered as a functional describing a good deal bound.
Therefore they appear to treat a special class of convex risk measures
but Theorem~\ref{thm2} shows that it is the only class 
giving good deal bounds.
The representation of GDV as $\rho_A$ is important in  
that it implies robustness of GDV to quantitative specification of
 investor's risk preference.
Notice however that $\rho_A$ is not necessarily normalized.
As long as treating $\rho_A$, the condition defining GDV is equivalent to
the no-cashout condition (NC) introduced in \cite{S04}:
$\rho_A(-x)\geq-\rho^0(x)$ for any $x\in L$.
In fact for any $x \in L$,
\begin{eqnarray*}
\rho^0(x)
&=&   \inf\{c\in\mathbb{R}|\mbox{ there exists  $m\in M$ such that }
      c+m+x\in L_+\} \\
&\geq&\inf\{c\in\mathbb{R}|\mbox{ there exists  $m\in M$ such that }
      c+m+x\in A\} \\
&=&   \rho_A(x),
\end{eqnarray*}
that is, the upper estimate for $\rho_A(-x)$ holds automatically.
The convexity of $\rho_A$ implies that NC is equivalent to $\rho_A(0)=0$.
Theorem~6.1(0th FTAP) of \cite{S04} states, in a more abstract setting,
a condition under which $\rho_A(0)=0$.
\fin
\end{rem}

\noindent
As mentioned in Introduction,
many papers (\cite{ES10}, \cite{KS07-2}, \cite{OS09}, \cite{X06},...)
treated risk indifference prices and some of them showed that
a risk indifference price yields a good deal bound.
On the other hand, Theorem~\ref{thm2} showed that a GDV is always 
a risk indifference price.
It therefore supports the  use of the operator $I$ in constructing
a good deal bound. 
We utilized however that a GDV has the Fatou property by definition.
It should be noted that $I(\rho)$ does not necessarily
have the Fatou property even if $\rho \in \calR$.
In other words, the operation does not necessarily preserve the Fatou
property (see Example~\ref{fatou} below).
Now we remark that it preserves the Lebesgue property that
also could be regarded as a natural continuity requirement for 
good deal bounds as well as the Fatou property.

\begin{prop}\label{Lebp1}
Let $\rho$ be a finite convex risk measure with 
the Lebesgue property and suppose that
there exists $Q^0 \in \calQ$ such that $\rho^\ast(Q^0) < \infty$.
Then, $I(\rho)$ is a finite GDV with the Lebesgue property.
\end{prop}

\proof
By Theorem \ref{Leb} and the existence of $Q^0\in\calQ$
such that $\rho^\ast(Q^0)<\infty$, we have, for any $x\in L$ and $m\in M$,
\begin{eqnarray*}
\rho(x+m)&=&   \max_{Q\in\calP}\{E_Q[-x-m]-\rho^\ast(Q)\}
               \geq E_{Q^0}[-x-m]-\rho^\ast(Q^0) \\
         &\geq&E_{Q^0}[-x]-\rho^\ast(Q^0)>-\infty.
\end{eqnarray*}
Therefore $I(\rho)$ is $(-\infty,\infty]$-valued by (\ref{eq-Ieta}), and so
it is a convex risk measure by Lemma~\ref{implicit}.
Since $\rho$ is finite, so is $I(\rho)$ by (\ref{eq-Ieta}).
Moreover for any $m\in M$, we have
\begin{equation}\label{irm}
I(\rho)(-m)=   \inf_{m^\prime\in M}\rho(-m+m^\prime)
                 -\inf_{m^\prime\in M}\rho(m^\prime)
           \leq \inf_{m^\prime\in M}\rho(m^\prime)
                 -\inf_{m^\prime\in M}\rho(m^\prime)=0.
\end{equation}
Therefore by Theorem~\ref{thm2}, 
it only remains to show that $I(\rho)$ has the Fatou property.
By (\ref{repL}),
it suffices to see that $I(\rho)$ has the Lebesgue property.
Note that $I(\rho)(m)\geq0$ for any $m\in M$ by the convexity.
For any $\alpha>0$, $\epsilon > 0$ and a sequence of measurable sets $A_n$
with $P(A_n) \to 0$, we have that
\begin{equation}\label{An}
\begin{split}
0 \leq& \ I(\rho)(-\alpha1_{A_n})
=  \inf_{m\in M}\rho(m-\alpha1_{A_n})-\inf_{m\in M}\rho(m)
\\
\leq & (1-\epsilon)\inf_{m \in M}\rho(\frac{m}{1-\epsilon}) + 
\epsilon  \rho(-\frac{\alpha}{\epsilon} 1_{A_n}) - 
\inf_{m \in M}\rho(m) \\
\to & -\epsilon \inf_{m \in M}\rho(m)
\end{split}
\end{equation}
as $n\to\infty$ by the Lebesgue property of $\rho$.
Since $\epsilon$ is arbitrary, we conclude
the Lebesgue property of $I(\rho)$ by Theorem \ref{Leb}.
\fin

\begin{prop}
For a finite convex risk measure $\rho$, 
the following are equivalent:
\begin{enumerate}
\item $\rho$  is a GDV with the Lebesgue property.
\item there exists a convex risk measure $\eta$
      with the Lebesgue  property, $\rho = I(\eta)$.
\end{enumerate}
\end{prop}
\proof 
1$\Rightarrow$2: This is because $\rho = I(\rho)$ by
Theorem~\ref{thm2}.

\noindent
2$\Rightarrow$1: 
By Lemma~\ref{implicit}, we have $\inf_{m \in M}\eta(m) \in \mathbb{R}$,
and so
\begin{equation*}
I(\eta)(x) = \inf_{m\in M}\eta(x+m) - \inf_{m \in M}\eta(m).
\end{equation*}
In particular we have (\ref{irm}) and (\ref{An}) with $\eta$ instead of
$\rho$. By the finiteness of $\rho = I(\eta)$, Theorem~\ref{Leb} can be
applied to have the result.
\fin

\begin{ex}\label{fatou}
Consider $L = L^\infty(\mathbb{R},\mathcal{F},\mathbb{P})$, 
where $\mathbb{P}$ is a normal distribution on $\mathbb{R}$.
Let $Q \in \calP$ have a compact support and define a sequence 
$\{Q_n\} \subset \cal{P}$  by $Q_n(A) := Q(A-n)$ for $A \in \mathcal{F}$, 
$n \in \mathbb{N}$.
Since $\{g \in L^\ast_+| g(1)=1\}$ is weak-* compact,
there exists a cluster point $\mu$ of $\{Q_n\}$.
Since $\{Q_n\}$ is not tight, $\mu \notin \mathcal{P}$.
Consider $M = \{x \in L | \mu(x) \leq 0\}$.
Observe that $\ol{L}^\ast = \{\mu\}$.
In fact if there exists $\nu \in \ol{L}^\ast$ and $x \in L$ with 
$\nu(x) > \mu(x)$, then $y:=x -\mu(x) \in M$ and $\nu(y) > 0$,
which is a contradiction.
Now consider $\rho \in \cal{R}$ defined as 
$\rho(-x) = \sup_{Q \in \calP}\mathbb{E}_Q[x]$.
Let us show that
$\rho^\ast(\mu)= 0$.
By $\rho(0) = 0$ we have  $\rho^\ast(\mu)\geq 0$ and
$\rho^\ast(Q_n)= 0$.
If $\rho(\mu) > 0$, then there exists $x \in L$ such that
$\mu(x) > \sup_{Q \in \calP}\mathbb{E}_Q[x]$, which contradicts that
$\mu$ is a cluster point of $Q_n$.
By the same reason, we have also that for any $m \in M$ and $x \in L$,
$\rho(m+x) \geq \mu(-m - x) \geq - \mu(x)$, so that
$I(\rho)$ is finite.
By  Lemma~\ref{implicit},
$I(\rho)^\ast(g)=\infty$ for any $g \in L^\ast_+ \setminus \ol{L}^\ast$, so
by Theorem~\ref{repfa}, we have $I(\rho)(-x) = \mu(x)$ for all $x \in L$.
To see that $I(\rho)$ does not have the Fatou property, 
consider the increasing sequence $x_n := 1_{(-\infty,n)}$.
Then $I(\rho)(-x_n) = 0$ while $I(\rho)(-x_\infty) = 1$.
\fin
\end{ex}

\subsection{Shortfall risk measures}
Here we treat shortfall risk measure as an application.
We presume an investor who sells a claim $x$.
When she sells $x$ with price $c$ and selects $m\in M$ as her strategy,
her final cash-flow is $c+m-x$,
and so its shortfall is  $(c+m-x)\wedge0$.
In general, shortfall risk is defined as a weighted expectation of the 
shortfall with a loss function.
A loss function is  a continuous strictly increasing convex function 
$l :\mathbb{R}_+ \to \mathbb{R}_+$ with $l(0)=0$.
This represents the seller's attitude towards risk.
To suppress the shortfall risk less than a certain level $\delta>0$
which she can endure, the least price she can accept is given as
\begin{equation}
\label{eqrhol}
\rho_l(-x):=\inf\{c\in\mathbb{R}|\mbox{ there exists }m\in M
            \mbox{ such that }E[l((c+m-x)^-)]\leq\delta\}.
\end{equation}
As shown in \cite{A11} and \cite{FS02}, $\rho_l$ is a convex risk measure
and it has the Fatou property under mild conditions.
However, it is not a GDV as $\rho_l(0)\neq 0$:

\begin{prop}
Any shortfall risk measure is not a GDV.
\end{prop}

\proof
For any shortfall risk measure $\rho_l$, (\ref{eqrhol}) implies that
\begin{eqnarray*}
\rho_l(0)
&=&   \inf\{c\in\mathbb{R}|\mbox{ there exists }m\in M
          \mbox{ such that }E[l((c+m)^-)]\leq\delta\} \\
&\leq&\inf\{c\in\mathbb{R}|l(c^-)\leq\delta\}
      =-l^{-1}(\delta)<0.
\end{eqnarray*}
Hence, $\rho_l\notin\calR$, from which $\rho_l$ is not a GDV.
\fin

\noindent
Now we show that a normalized shortfall risk measure can be a GDV.
Define $\wh{\rho_l}$ as $\wh{\rho_l}(x):=\rho_l(x)-\rho_l(0)$.

\begin{prop}
\label{prop-shortfall}
If $\wh{\rho_l}\in\calR$, then $\wh{\rho_l}$ is a GDV.
\end{prop}

\proof
In light of Theorem~\ref{thm2}, it suffices to see 
$I(\wh{\rho_l}) = \wh{\rho_l}$.
Since $\wh{\rho_l}(m) \geq - \wh{\rho_l}(-m) \geq 0$ for $m \in M$, 
we have $\inf_{m \in M}\wh{\rho_l}(m) = 0$, and so
$I(\wh{\rho_l})(x) = \inf_{m\in M}\rho_l(m+x) - \rho_l(0)$.
Now let us observe that $\inf_{m\in M}\rho_l(m+x)=\rho_l(x)$
for any $x\in L$.
$\inf_{m\in M}\rho_l(m+x)\leq\rho_l(x)$ holds clearly.
Fix $m\in M$ and $c>\rho_l(m+x)$ arbitrarily.
Then there exists $m^\prime\in M$ such that
$E[l((c+m^\prime+m+x)^-)]\leq\delta$.
Since $m^\prime+m\in M$, we have $c\geq\rho_l(x)$.
\fin

\setcounter{equation}{0}
\section{Relevant good deal valuations}
\subsection{Fundamental Theorem of Asset Pricing}
We have seen that the condition $\calQ \neq \emptyset$ is 
equivalent to the existence of a GDV. 
Example~\ref{ex-sec4} below shows that $\calQ \neq \emptyset$ is
not sufficient to rule out arbitrage opportunities in general.

\begin{ex}
\label{ex-sec4}
Let $A \in \calF$ with $P(A)\in(0,1)$, $m^\prime := 1_A$ and
$M=\{cm^\prime|c\geq0\}-L_+$.
Any probability measure $Q\in\calP$ with
$Q(A)=0$ is in $\calQ$.
On the other hand, $cm^\prime$ with $c>0$ brings an arbitrage opportunity.
\fin
\end{ex}

\noindent
Kreps~\cite{K81} showed that 
$\calQ^e \neq \emptyset$ is equivalent to NFL, that is,
$\ol{M} \cap L_+ = \{0\}$.
Here we prove that
$\calQ^e \neq \emptyset$ is equivalent to the existence of a relevant GDV,
that is, a relevant convex risk measure which is a GDV.

\begin{thm}[FTAP]
\label{thm3}
The following are equivalent:
\begin{enumerate}
\item $\calQ^e \neq \emptyset$.
\item $\ol{M}\cap L_+=\{0\}$.
\item There exists a relevant GDV.
\end{enumerate}
\end{thm}

\proof
2$\Rightarrow$1:
For any $a,b \in \mathbb{R}$, the set $\{x \in L| a \leq x \leq b  \}$
 is compact in $\sigma(L,L^\dagger)$.
In fact if $L = L^\infty$, then $L^\dagger = L^1$ and 
 $\sigma(L,L^\dagger)$ is the weak-* topology.
The compactness then follows from the Banach-Alaoglu theorem.
It suffices then to notice that $L^\infty \subset L$, $L^\dagger \subset L^1$
as sets of random variables  and the natural inclusion 
$(L^\infty, \sigma(L^\infty,L^1))\to (L,\sigma(L,L^\dagger))$ is continuous.
Therefore we can prove the existence of an element of $\calQ^e$ 
in exactly the same manner as in the proof of Theorem 5.2.3 of \cite{DS06}.

\noindent
1$\Rightarrow$3: This is from Lemma~\ref{lem0-2}.

\noindent
3$\Rightarrow$2:
Let $\rho$ be a relevant GDV.
We have
$\rho(x)=\sup_{Q\in\calQ}\{\mathbb{E}_Q[-x]-c(Q)\}$ by Item 3 of
Theorem~\ref{thm2}.
Since $\rho(-z) > 0$ for all $z \in L_+$ by the relevance, it
suffices to see that $\rho(-\ol{m}) \leq 0$
for any $\ol{m} \in\ol{M}$.
If there exists $\ol{m} \in \ol{M}$ with $\rho(-\ol{m}) > 0$, 
there exists $Q \in \calQ$ such that $\mathbb{E}_Q[\ol{m}] > c(Q) \geq
\sup_{m \in M} \mathbb{E}_Q[m]$.
The last inequality is from the fact that 
$\rho(-m) \leq 0$ for all $m\in M$.
This contradicts that $\ol{m}$ is in the closure of $M$
in $\sigma(L, L^\dagger)$.
\fin

\noindent
Now we give a set of 
equivalent conditions for GDV to be relevant.
Let
\begin{eqnarray}
\label{eq-Mrho}
M^\rho&:=&\{x-\rho(-x)|x\in L, \rho(-x) < \infty \}-
L_+=\{x\in L|\rho(-x)=0\}-L_+ \nonumber \\
      &=& \{x\in L|\rho(-x)\leq0\}.
\end{eqnarray}
Note that $M^\rho$ is a convex set including $M$ and interpreted as 
the set of the $0$-attainable claims of an extended market where
an investor offers prices for all $x \in L$ by using $\rho$ as her
pricing functional.
In light of Theorem~\ref{BF}, $M^\rho$ is closed in $\sigma(L,L^\dagger)$.
Therefore NFL for this extended market is $M^\rho \cap L_+ = \{0\}$.

\begin{thm}
\label{thm4}
For a GDV $\rho$, the following are equivalent:
\begin{enumerate}
\item $\rho$ is relevant.
\item $-\wh{\rho^0}(x-z) < \rho(-x)$ for any $x \in L$ and 
       $z \in L_+\setminus \{0\}$.
\item $-\rho^0(x-z) < \rho(-x)$ for any $x \in L$ and 
       $z \in L_+\setminus \{0\}$.
\item  $M^\rho \cap L_+ = \{0\}$.
\end{enumerate}
\end{thm}

\proof
1$\Rightarrow$2: 
By the relevance and Theorem~\ref{thm2},
for any $z \in L_+\setminus \{0\}$, there exists $Q(z) \in \calQ$ such that
$\mathbb{E}_{Q(z)}[z] > \rho^\ast(Q(z))$. Therefore,
\begin{equation*}
-\wh{\rho^0}(x-z) = \inf_{Q \in \calQ}\mathbb{E}[x-z] \\
\leq \mathbb{E}_{Q(z)}[x-z] < \mathbb{E}_{Q(z)}[x]-\rho^\ast(Q(z)) 
\leq \rho(-x).
\end{equation*}

\noindent
2$\Rightarrow$3: This is from Lemma~\ref{lem0-2}.

\noindent
3$\Rightarrow$1: For a given $z \in L_+\setminus \{0\}$, let $x=z$. 

\noindent
1$\Rightarrow$4: This is because 
$\rho$ separates $M^\rho$ and $L_+\setminus \{0\}$.

\noindent
4$\Rightarrow$1: If $\rho$ is not relevant, then there exists 
$z \in L_+\setminus \{0\}$ such that $\rho(-z) =0$. 
In particular $z \in M^\rho$,  which is a contradiction.
\fin

\begin{rem}\label{staumrem}
Item~3 of Theorem~\ref{thm4} is the no-near-arbitrage condition (NNA)
introduced in \cite{S04}.
Theorem~6.2 of \cite{S04} states a condition under which 
$\rho_A$ satisfies NNA.
Proposition~\ref{cor5} below may be regarded as  its counterpart.
\fin
\end{rem}

\begin{prop}
\label{cor5}
Let $\rho$ be a GDV.
If there exists $Q_0 \in\calQ^e$ such that
$\rho^\ast(Q_0)=0$, then $\rho$ is relevant.
The reverse implication holds true if $\rho$ is coherent.
\end{prop}

\proof
The relevance is clear from Theorem~\ref{BF}.
The converse is the Halmos-Savage theorem (see e.g. \cite{Delrel}).
\fin

\noindent
Note that for $Q \in \calP$ and $\rho\in\calR$,
$\rho^\ast(Q)=0$ is equivalent to that
$-\rho(x) \leq \mathbb{E}_Q[x] \leq \rho(-x)$ for all $x \in L$.
Therefore such $Q$ is interpreted as a consistent pricing kernel of the
extended market $M^\rho$.
The following example shows that the coherence in 
the second assertion of Proposition~\ref{cor5} cannot be dropped.
In other words, there is no strictly positive consistent pricing kernel
in general even if $M^\rho$ satisfies NFL: $M^\rho \cap L_+ = \{0\}$.

\begin{ex}\label{exnoncoh}
Set $\Omega=\{\omega_1, \omega_2\}$, and $M=L_-$.
Denoting $q:=Q(\{\omega_1\})$, we can identify $q$ with $Q \in \calQ$.
From this viewpoint, $\calQ$ and $\calQ^e$ are corresponding to
$[0,1]$ and $(0,1)$ respectively.
Consider $\rho(-x)=\sup_{Q\in\calQ}\{\mathbb{E}_Q[x]-c(Q)\}$
with   $c(Q)=q^2$. Then we have $\rho^\ast(Q)=c(Q)$.
Denoting $z_i:=z(\omega_i)$ for $i=1,2$, we have
$\rho(-z)=\sup_{Q\in\calQ}\{\mathbb{E}_Q[z]-c(Q)\}
         =\sup_{q\in[0,1]}\{qz_1+(1-q)z_2-q^2\}
         =\sup_{q\in(0,1)}\{qz_1+(1-q)z_2-q^2\}>0$
for any  $z\in L_+\setminus\{0\}$.
Thus, $\rho$ is a noncoherent relevant GDV.
On the other hand, there is no $q\in(0,1)$ with $c(Q)=0$.
\fin
\end{ex}

\begin{rem}\label{bionnadalrem}
In \cite{B09}, NFL refers to the condition that
\begin{equation*}
\overline{\mathrm{cone}(M^\rho)} \cap L_+ =\{0\},
\end{equation*}
where $\overline{\mathrm{cone}(M^\rho)}$ is the closure of
$\mathrm{cone}(M^\rho) = \{\lambda m; m \in M^\rho, \lambda  \geq 0\}$ 
in $\sigma(L, L^\dagger)$, which is a different condition to 
$M^\rho \cap  L_+ = \{0\}$ unless $\rho$ is coherent.
This alternative  definition of NFL enabled to establish the 
equivalence between NFL of $\rho$ and the existence of 
$Q_0 \in \calQ^e$ with $\rho^\ast(Q_0)=0$ in \cite{B09}.
In fact since
$\overline{\mathrm{cone}(M^\rho)}$ becomes a cone,
the same argument as the proof of 2$\Rightarrow$1 of Theorem~\ref{thm3}
can apply to have $Q_0 \in \calQ^e$ with $\mathbb{E}_{Q_0}[m] \leq 0$
for all $m \in M^\rho$.
Since $x - \rho(-x) \in M^\rho$ for all $x \in L$, we have
$\rho^\ast(Q_0)=0$.
Note however that
$\mathrm{cone}(M^\rho)$ does not have any interpretation as 
the set of the 0-attainable claims in general. For instance, 
in the model of the preceding example,
we can find $x\in L$ with $\rho(-x)\leq0$ and $\lambda > 0$ satisfying
$\rho(-\lambda x) > 0$.
Therefore, it seems not adequate, from economical point of view,
to adapt such a definition of NFL.
Consequently, the existence of $Q_0$ with $\rho^\ast(Q_0)=0$
may not be considered as a necessary condition for $\rho$ to be a 
reasonable pricing functional.
\end{rem}

\subsection{When are all good deal valuations relevant?}
As seen in Theorem \ref{thm4}, when we extend the underlying market $M$ to
$M^\rho$ by using a GDV $\rho$ as pricing functional,
the extended market $M^\rho$ remains to satisfy NFL if and only if
$\rho$ is relevant.
Therefore markets in which any GDV is relevant are stable
against such extensions of the market.
Here we study necessary and (or) sufficient conditions
under which all (coherent) GDVs are relevant.

\begin{thm}
\label{thm5}
Suppose $\calQ^e\neq \emptyset$ and consider the following conditions:
\begin{enumerate}
\item Any GDV is relevant.
\item $\wh{\rho^0}(z)<0$ for any $z\in L_+\setminus\{0\}$.
\item $\calQ=\calQ^e$.
\item[3$^\prime$] 
         $\calQ=\calQ^e$ and $\calQ^e$ is $\sigma(L^\dagger,L)$-compact.
\item Any coherent GDV is relevant.
\end{enumerate}
Then, we have 1$\Leftrightarrow$2, 2$\Rightarrow$3, 3$^\prime\Rightarrow$2,
3$\Leftrightarrow$4.
\end{thm}

\proof
1$\Rightarrow$2:
Assume that there exists  $z_0 \in L_+\setminus\{0\}$
such that $\wh{\rho^0}(z_0)=0$.
Then $\inf_{Q \in \calQ}\mathbb{E}_Q[z_0] = 0$,
so that we can define $\rho \in \calR$ as 
\begin{equation*}
\rho(-x) = \sup_{Q \in \calQ}\{\mathbb{E}_Q[x]-\mathbb{E}_Q[z_0]\}.
\end{equation*}
This is a GDV by Theorem~\ref{thm2} but not relevant.
In fact $\rho(-z_0)=0$.

\noindent
2$\Rightarrow$1:
Let $\rho$ be a GDV. Then by Item~5 of Theorem~\ref{thm2}, 
$\rho(-z) \geq -\wh{\rho^0}(z) > 0$ for any $z \in L_+\setminus\{0\}$.

\noindent
2$\Rightarrow$3:
If $\calQ\neq\calQ^e$, then there exists $Q^*\in\calQ\backslash\calQ^e$.
Denoting $A=\{\mathrm{d}Q^*/\mathrm{d}\mathbb{P}>0\}$,
$\wh{\rho^0}(1_{A^c})=    \sup_{Q\in\calQ}\mathbb{E}_Q[-1_{A^c}]
                     \geq E_{Q^*}[-1_{A^c}]
                     =0$, 
while $1_{A^c} \in L_+\setminus\{0\}$.

\noindent
3$^\prime\Rightarrow$2:
By compactness we have for any $z\in L_+\setminus\{0\}$,
\[
\wh{\rho^0}(z)=\sup_{Q\in\calQ}\mathbb{E}_Q[-z]
              =\sup_{Q\in\calQ^e}\mathbb{E}_Q[-z]
              =\max_{Q\in\calQ^e}\mathbb{E}_Q[-z]
              <0.
\]

\noindent
3$\Rightarrow$4:
Any coherent GDV $\rho$ is represented as
$\rho(x)=\sup_{Q\in\wh{\calQ} }\mathbb{E}_Q[-x]$,
for some convex set $\wh{\calQ}\subset\calQ = \calQ^e$.
Therefore $\rho$ is relevant.

\noindent
4$\Rightarrow$3:
If $\calQ\neq\calQ^e$ then we can
take $Q^*$ and  $A$ in the same way as ``2$\Rightarrow$3".
Let
$\rho(x)=\sup_{Q\in\calQ, Q(A)=1}\mathbb{E}_Q[-x]$.
Then $\rho \in \calR$ since $Q^\ast(A)=1$.
By Theorem~\ref{thm2}, $\rho$ is a coherent GDV but not relevant
since $\rho(1_{A^c})=0$.
\fin

\noindent
The implications ``3$\Rightarrow$3$^\prime$",
``3$\Rightarrow$1 (or 2)" and ``2$\Rightarrow$3$^\prime$" in Theorem \ref{thm5}
do not hold in general.
We illustrate counterexamples.

\begin{ex}
\label{ex5-1}
We give an example satisfying Item~3 of Theorem~\ref{thm5}
which does not satisfy Items 1 nor $3^\prime$.
Set $\Omega=\mathbb{R}$, $L=L^\infty$ and 
$\mathbb{P}(\mathrm{d}u)=\phi(u)\mathrm{d}u$,
where $\phi(u)$ is the standard normal density.
We consider the set of the mixed normal distributions.
Let $V$ be the set of all probability measures on $(0,\infty)$, 
\begin{equation*}
Q_\mu(\mathrm{d}u):=\int\frac{1}{\sqrt{v}}\phi(u/\sqrt{v})\mu(\mathrm{d}v)
                    \mathrm{d}u
\end{equation*}
for $\mu \in V$, and $\wh{\calQ} := \{Q_{\mu}| \mu \in V\}$.
Define $M$ as 
\[
M=\{m\in L^\infty| \mathbb{E}_Q[m]\leq0\mbox{ for any }Q\in\wh{\calQ}\}.
\]
Note that all bounded odd functions are in $M$ and
$\wh{\calQ}\subset \calQ^e \subset\calQ$.
Now we show that $\wh{\calQ}$ is $\sigma(L^1,L^\infty)$-closed.
Let $\{\mu_n\} \subset V$ be a sequence with $Q_{\mu_n}\to Q$ 
in $\sigma(L^1, L^\infty)$.
Denote $y_w(u):=e^{iwu}$ for any $w$, $u\in\mathbb{R}$, where $i=\sqrt{-1}$.
We have
\[
E_{Q_{\mu_n}}[y_w]=\int e^{-\frac{v}{2}w^2}\mu_n(dv),
\]
which has the form of the Laplace transform of $\mu_n$.
Since $E_{Q_{\mu_n}}[y_w]\to \mathbb{E}_Q[y_w]$ and
$\lim_{w \to 0}\mathbb{E}_Q[y_w]=1$, 
the continuity theorem of Laplace transforms (see Theorem~XIII.1.2 of
\cite{Feller}) implies the existence of $\mu \in V$ such that 
\begin{equation*}
\mathbb{E}_Q[y_w]=\int e^{-\frac{v}{2}w^2}\mu(\mathrm{d}v),
\end{equation*}
which is the characteristic function of an element of $\wh{\calQ}$.
Hence, $Q \in \wh{\calQ}$.

Note that $\wh{\calQ}=\calQ^e=\calQ$. 
In fact if there exists $Q^\ast \in  \calQ \setminus \wh\calQ$, 
by the Hahn-Banach theorem there exists $x \in L$ such that
$\mathbb{E}_{Q^\ast }[x] > \sup_{Q \in \hat{\calQ}}\mathbb{E}_Q[x] =: \alpha$.
However, $x -\alpha \in M$ and $\mathbb{E}_{Q^\ast}[x-\alpha] > 0$,
which contradicts $Q^\ast \in \calQ$.
On the other hand, $\calQ$ is not compact.
In fact for the sequence  $\mu_n:=\delta_{1/n}$ for $n\in \mathbb{N}$,
where $\delta_u$ is the Delta measure concentrated on $\{u\}$, 
$\{Q_{\mu_n}\}$ does not have a cluster point in $\wh{\calQ}$.

Finally, we construct a GDV $\rho$ which is not relevant.
Letting $y(u):=u^2$, we define $\rho$ as 
$\rho(-x) = \sup_{Q \in \calQ}\{\mathbb{E}_Q[x]-c(Q)\}$ with
$c(Q)=\mathbb{E}_Q[y]$.
Obviously, we have $\rho(0)=0$ and $\rho(-y)=0$.
\fin
\end{ex}

\begin{ex}
Here we see that the implication ``2$\Rightarrow$3$^\prime$" in
Theorem~\ref{thm5} does not hold.
We modify Example \ref{ex5-1} as follows.
Let $\mu_0 \in V$ be fixed and
$\wh{\calQ}_0 := \{Q_{\nu}| \nu = (\mu_0 + \mu)/2, \mu \in V\}$.
By the same argument as in Example~\ref{ex5-1}, 
we can prove the closedness and noncompactness of $\wh{\calQ}_0$ and
that $\wh{\calQ}_0 = \calQ = \calQ^e$.
This model however satisfies Item~2 of Theorem \ref{thm5} since 
\begin{equation*}
\wh{\rho^0}(z)
=   \sup_{Q\in\wh{\calQ}_0 }\mathbb{E}_Q[-z]
=   \frac{1}{2}E_{Q_{\mu_0}}[-z]+\frac{1}{2}\sup_{\mu \in V}E_{Q_\mu}[-z]
\leq\frac{1}{2}E_{Q_{\mu_0}}[-z] < 0.
\end{equation*}
\fin
\end{ex}

\noindent
We conclude the paper with  one more example, 
which is a simple model taking transaction cost into account.
In the following example, a model satisfying Item~3$^\prime$ of 
Theorem~\ref{thm5} is constructed.

\begin{ex}
Let $\Omega = \{\omega_0,\omega_1,\dots,\omega_n\}$ and 
the Arrow-Debreu securities for the $n$ states 
$\omega_1,\dots \omega_n$ be tradable in a market subject to bid-ask spread.
Denote by $a_{1,j}$, $a_{-1,j}$ the ask and bid prices for the state $\omega_j$
respectively for each $j=1,\dots, n$.
Let $D := \{-1,1\}^n$.
If $a_{-1,j} \geq 0$ for each $j$ and $\sum_ja_{1,j} \leq 1$,
then for any $d \in D$, a probability measure $Q_d$ on $\Omega$
is uniquely determined by $Q_d(\{\omega_j\}) = a_{d(j),j}$ for $j=1,\dots, n$,
and $Q_d(\{\omega_0\}) = 1 - \sum_{j=1}^na_{d(j),j}$.

Now let 
\begin{equation*}
M = \{x \in L| \mathbb{E}_d[x]\leq 0 \text{ for all } d \in D\}
  = \l\{x - \max_{d \in D}\mathbb{E}_d[x]| x \in L\r\} - L_+,
\end{equation*}
where $\mathbb{E}_d$ is the expectation under $Q_d$. 
Note that any cash-flow $x \in L$ can be uniquely represented as a sum of
a constant and the Arrow Debreu securities and that
the price for replicating $x$ is 
$\max_{d \in D}\mathbb{E}_d[x]$. 
Therefore $M$ is actually  the set of the $0$-attainable claims in this market.
By the same separation argument as in the preceding examples,
we can show
\begin{equation*}
\calQ = \l\{\sum_{d \in D} \lambda_d Q_d | \lambda_d \geq 0 \text{ for all }
        d \in D \text{ and } \sum_{d \in D}\lambda_d = 1\r\}.
\end{equation*}
This set is compact because the set of $(\lambda_d)$ is a
finite dimensional simplex.
If $\sum_ja_{1,j} < 1$ in addition, then $\calQ = \calQ^e$ and so,
Item~3$^\prime$ of Theorem~\ref{thm5} is satisfied.
Consequently, any GDV in this market is relevant.
Remark that $\sum_ja_{1,j} < 1$ is a  condition which requires
market makers not to offer a set of prices which leads 
an apparent arbitrage opportunity for themselves.
\fin
\end{ex}

\begin{center}
{\bf Acknowledgements}
\end{center}
The authors would like to thank Professors Freddy Delbaen and Martin Schweizer
for their valuable comments and suggestions.
This work was done when the authors were Visiting Professors of ETH Zurich.
Takuji Arai was supported by Scientific Research (C) No.22540149
from the Ministry of Education, Culture, Sports, Science and Technology of
Japan.
Masaaki Fukasawa was supported by Japan Science and Technology Agency,
CREST.


\end{document}